# New process for high optical quality InAs quantum dots grown on patterned GaAs(001) substrates

Pablo Alonso-Gonzalez1,3, Luisa González1, Yolanda González1, David Fuster, Iván Fernández-Martínez1, Javier Martín-Sanchez1and Leon Abelmann2

*1 Instituto de Microelectrónica de Madrid (CNM, CSIC), Isaac Newton 8, 28760, Tres Cantos, Madrid, Spain*

*2 SMI, MESA+Institute of Nanotechnology, University of Twente, PO Box 217, 7500 AE, Enschede, The Netherlands*

Abstract

This work presents a selective ultraviolet (UV)-ozone oxidation-chemical etching process that has been used, in combination with laser interference lithography (LIL), for the preparation of GaAs patterned substrates. Further molecular beam epitaxy (MBE) growth of InAs results in ordered InAs/GaAs quantum dot (QD) arrays with high optical quality from the first layer of QDs formed on the patterned substrate. The main result is the development of a patterning technology that allows the engineering of customized geometrical displays of QDs with the same optical quality as those formed spontaneously on flat non-patterned substrates.

1. Introduction

Ordered quantum dot (QD) configurations are demanded for future new nano-electronics and quantum computation technologies [1]. In particular, for the development of quantum devices, precise site control and high crystalline quality of the nanostructures is necessary [2].

For this purpose, many approximations have been studied to overcome the randomness of a spontaneously selfassembled nanostructure [3] while at the same time keeping the nanostructure defect-free, which is its main advantage. To achieve this aim, a quite widespread strategy is the modification of the surface to create preferential nucleation sites for the nanostructures by using patterned substrates. In this way, highly ordered arrays of semiconductor QDs have already been obtained by using different lithographic techniques [4–9].

The main drawback of these approaches is the loss of crystalline perfection and the incorporation of impurities in the nanostructures associated with the technological processes involved in the patterning. In this situation, the QDs formed at the patterned surface show poor optical properties. Thus, in order to achieve high photoluminescence (PL) efficiency, similar to that obtained from self-assembled QDs on unprocessed substrates, it is necessary to separate the QDs far from the patterned interface. Using this method, the growth of stacks of several QD layers on the patterned substrates [8] has been demonstrated as a possible solution for obtaining simultaneously ordered QDs and high optical quality. However, despite the progress made in minimizing undesired effects related to the technology used for patterning, it is still necessary to develop new technological processes that ideally would allow for the fabrication of QDs with accurate control of size, shape and position while keeping them as defect-free nanostructures.

In this work we present results that demonstrate a way to obtain large-area ordering of QDs with high optical quality on a patterned substrate. This is obtained from the first QD layer without the need for using stacks of QD layers or separating the QD layer from the patterned surface by thick buffer layers that smooth the patterned profile. Our approach consists of using a combination of a selective UV-ozone cleaning/oxidation process and chemical etching to produce configurations of ordered concave pits, or convex humps, on a previously masked GaAs substrate. The process followed results on a bare GaAs surface patterned with nanoholes ready for further epitaxial growth of InAs QDs in a molecular beam epitaxy (MBE) system. The final result is the formation of ordered QDs over a short distance (6.8 nm) of the patterned substrate with emission properties similar to those from selfassembled QDs.

2. Experimental details

Our experiments start with a 500 nm thick undoped epitaxial GaAs layer grown by MBE on a commercial epi-ready GaAs(001) substrate (from this point onwards, epitaxial substrate). On top of the epitaxial substrate, a direct negative resist (MaN2403, MicroResist Technology, Berlin, Germany) was spin-coated. The initial photoresist thickness was 215 nm, which reduces somewhat after bake-out. The exposure of the photoresist was carried out with a laser interference setup [10]. In this setup a 266 nm quadrupled Nd:YAG laser beam is projected onto a Lloyd's mirror and the resulting periodic fringe interference pattern with a periodicity of 300 nm is used for the exposure. By rotating the sample at an angle of 45◦ and performing a second exposure, we obtain after development an array of holes aligned in a square pattern.

The next step is to transfer this pattern fabricated by LIL to the GaAs substrate. For that purpose we first oxidize the bare GaAs nanoholes by exposing the sample to ozone [11, 12]. The oxidation process consists of exposure of the GaAs substrate to UV light with a wavelength of $\lambda$ = 184.9 nm produced in a low-pressure Hg discharge lamp in open-air conditions. In this situation, atomic oxygen and ozone are simultaneously produced upon interaction of the UV light with the oxygen present in the air atmosphere [13]. The ozone reacts with the GaAs surface, producing $Ga_2O_3$, $As_2O_5$ and $As_2O_3$ [11]. After ozone exposure, the bottom of the holes is now oxidized GaAs, which can be removed chemically by dipping the sample in 1 M citric acid solution for 60 s. The combination of oxide formation/oxide dissolution leaves an atomically smooth GaAs surface, as characterized by atomic force microscopy (AFM). Previous calibrations of our experimental setup show that 1 nm of the GaAs substrate is removed per oxidation/oxide dissolution cycle, for an ozone exposition of 60 s. By repeating this process we can produce nanoholes in the GaAs substrate, tuning their depth in an extremely accurate way.

The next step to achieve a GaAs patterned substrate is to get rid of the resin existing between the nanoholes. The resin was selectively removed by an appropriate mix of developer MF440, remover mr-REM 660 and hydrofluoric acid (HF). Oxygen plasma was also used to guarantee surface cleanliness.

Once the patterned GaAs substrates were obtained, the samples were loaded into the MBE chamber. Prior to the growth of the InAs QDs, the surface has to be treated in order to remove native oxides and other possible residual contaminants. This process has to be carried out at a low enough substrate temperature, as close as possible to $T_s \leq 500$ ◦C, in order to avoid the degradation or smoothing of the pattern [14]. With this aim, different low-temperature treatments of the surface have been reported previously [15, 16]. In our case, the sample, introduced in the chamber together with a non-patterned GaAs epitaxial substrate as a reference, was treated by exposure to an atomic hydrogen flux with a pressure of $P(H_2)$ = 10-5 mbar at a substrate temperature of $T_s$ = 450 ◦C. During this process the As cell was kept open with a measured value of beam equivalent pressure of BEP $(As_4)$ = 2 × 10-6 mbar. The reflection high-energy electron diffraction (RHEED) diagram showed a clear 2× periodicity along the [110¯] direction on the reference sample after the first minute of exposure. The process is maintained for 20 min in order to eliminate possible residual organic contaminants from the resin on the patterned sample.

Trying to keep most of the growth process at $T_s$ < 500 ◦C, after the oxide removal process, a 6.8 nm thick GaAs buffer layer was grown at $T_s$ = 450 ◦C using the atomic layer molecular beam epitaxy (ALMBE) technique [17].

The formation of InAs QDs was carried out by depositing InAs at $T_s$ = 510 °C up to the critical thickness ($\theta_c$ = 1.7 ML), as observed by a two-dimensional to threedimensional (2D–3D) change in the RHEED diagram of the reference sample (without pattern). InAs was grown following a growth sequence consisting of 0.1 monolayers (ML) of InAs deposition at a growth rate of 0.05 ML s-1 followed by a pause of 2 s under an As2 flux. The different parameters involved in this epitaxial growth were optimized to ensure atomic flatness of the interface and the highest optical quality of the QDs [18].

For PL investigation, we have covered the InAs QD with a 15 nm thick GaAs layer. For AFM characterization, after the GaAs cap layer, we again deposited 1.7 ML of InAs for QD formation on the surface.

3. Results

Figure 1(a) shows the AFM image of a typical hole array as starting configuration. In this case, the diameter of the circular windows is 220 ± 20 nm and their average depth is 200 nm, which corresponds to the previously deposited resist thickness. The centre-to-centre distance between contiguous holes is 300 ± 10 nm, which is in agreement with the settings in the LIL setup.

Figure 1(b) shows an AFM image of the GaAs surface after four oxidation/oxide dissolution cycles. The average diameter of the obtained patterned holes is 280±33 nm and the depth is 4 ± 0.5 nm. Comparing with the original apertures on the resin layer, the final holes transferred to the GaAs substrate are wider while the distance between holes is totally preserved (290 ± 10 nm). The surface between the holes is very smooth, with a peak-to-peak roughness of 0.6 nm. With respect to the morphology at the bottom of the holes, it presents a higher, although still low, roughness of 1 nm.

Figure 2 shows the comparative AFM image of the superficial InAs QDs grown on patterned and non-patterned substrates (figures 2(a) and (b), respectively) for samples with buried QDs (for PL characterization) and QDs at the surface. On the patterned substrate, figure 2(a), a square configuration of ordered QDs is clearly observed. The spatial distribution of the buried QDs seems to be perfectly replicated onto the QDs grown at the surface following the initial pattern ordering (figure 1(b)). On the other hand, the AFM image of the reference sample (figure 2(b)) shows a typical random nucleation of QDs.

We have also observed that, due to imperfections in the LIL process, some peripheral parts of the initial pattern present an eventual overlapping of the holes along the [110⁻] direction (figure 3(a)). Related to this geometry of the initial pattern, linear configurations of QDs were obtained (figure 3(b)). The features of the pattern will probably be magnified during the 6.8 nm thick GaAs buffer layer growth due to the enhanced Ga diffusion along [110⁻].

The average base diameter and height of QDs in both square and linear configurations are 69 ± 10 and 18 ± 5 nm, respectively. The QDs on the reference sample present a similar diameter of 67 ± 10 nm and height of 17 ± 5 nm. The densities of QDs in the patterned and reference samples are 1.9 × 109 and 2.5 × 109 cm-2, respectively.

Figure 4 shows the normalized 20 K PL spectra for InAs QDs grown on the patterned substrate (dotted line) and on the reference sample (continuous line). The QD distribution of these samples is shown on figure 2. Both the emission energy and the line width of the main PL peak ($E_p$ = 1.2, 1.18 eV, FWHM = 43, 44 meV for the patterned and reference samples, respectively) are very similar. In the ordered QD PL spectrum we observe another peak of lower intensity at lower energies. We exclude that these two PL peaks correspond to the ground and excited states of the QDs, as the relative intensity remains constant with the excitation power. The appearance of two PL peaks might be related to a bimodal size distribution in the QDs [19] of the patterned sample. Nevertheless, it remains unclear why the low-energy PL peak is not observed in the reference sample, taking into account that the size distributions of QDs on both patterned and non-patterned substrates are very similar.

These results show that, after the whole technological process for patterning the GaAs substrates and further growth of an extremely thin GaAs buffer layer (6.8 nm), we obtained a patterned substrate with the same characteristics of flatness and cleanliness of an unprocessed substrate. Instead of the random distribution of self-assembled QDs on a flat surface, the patterned substrates impose an ordering of the QD layout. In turn, improvements in size uniformity in the initial patterning could be critical for the degree of ordering achieved in the final QD distributions.

4. Summary

In summary, in this work we have demonstrated that ordered QDs with the same optical quality as shown by random selfassembled QDs can be obtained without the need for stacking a number of QD layers to separate the active QDs far from the patterned substrate. The fabrication process that has been developed is based on a novel combination of ozone oxidation/oxide etching processes on a previously LIL-masked substrate. The oxides that were formed were dissolved in citric acid solution. Repeating the oxidation/oxide dissolution process, the mask pattern is transferred to the substrate, leaving arrays of nanoholes with a depth that can be tuned up as desired. Upon further epitaxial growth, ordered QD distributions have been obtained with PL emission similar to that obtained on self-assembled QDs grown on unprocessed substrates.

Demonstrated for the case of patterning by LIL, this process can be extended to any other lithographic technique.

Acknowledgements

The authors gratefully acknowledge financial support from the Spanish MEC and CAM through projects nos TEC- 2005-05781-C03-01, NAN2004-09109-C04-01, ConsoliderIngenio 2010 CSD2006-0019 and S-505/ESP/000200, and from the European Commission through the SANDIE Network of Excellence (no. NMP4-CT-2004-500101). PAG and JMS thank the I3P program. Mr H Kelderman of the MESA+research institute is gratefully acknowledged for production of the LIL patterns.

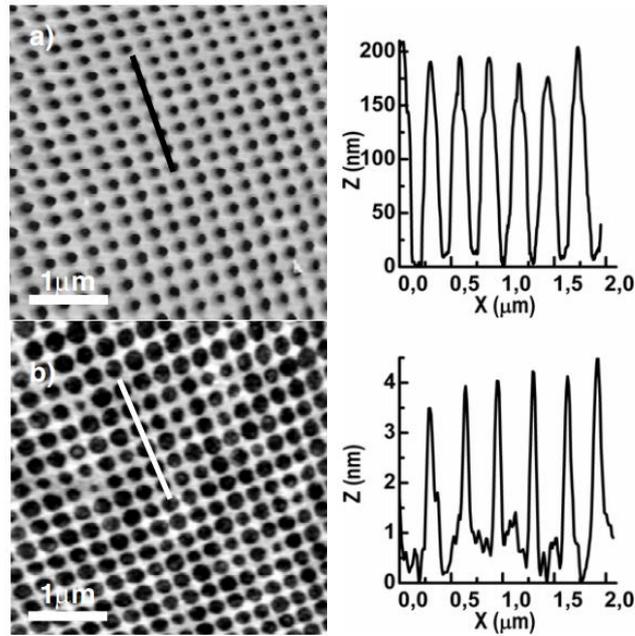

Figure 1. 5 × 5 $\mu m^2$ atomic force microscopy (AFM) images of (a) the initial patterned surface fabricated by laser interference lithography (LIL) with opened nanoholes over a resin layer and (b) the GaAs etched surface after four oxidation/oxide dissolution cycles. A profile along the line drawn on the respective figures is shown on the right.

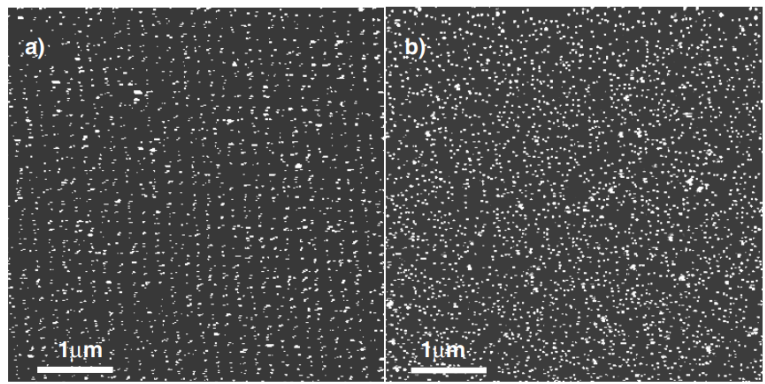

Figure 2. (a) 10 × 10 µm2 atomic force microscopy (AFM) image of (a) square superficial QD distribution grown on a selective etched GaAs substrate and (b) QDs grown on the non-patterned reference sample.

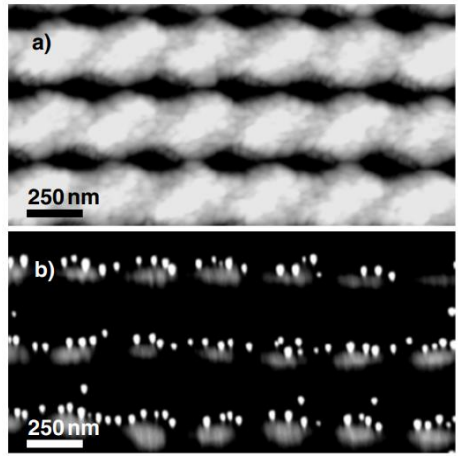

Figure 3. 2 × 1 µm² atomic force microscopy (AFM) images of (a) the starting resin pattern fabricated by laser interference lithography (LIL) showing overlapping of the holes along the [1-10] direction and (b) aligned QDs grown on the resulting GaAs patterned substrate.

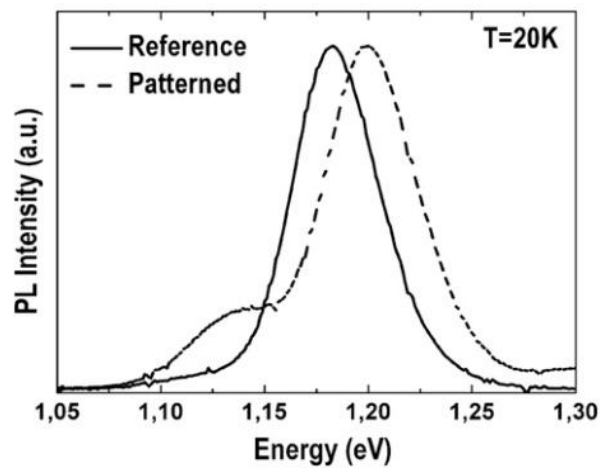

Figure 4. Comparison of the normalized 20 K photoluminescence (PL) spectra of the ordered InAs QD array (dotted line) and the simultaneously grown reference sample (continuous line). The emission energy and the line width of the main PL peak are $E_p$ = 1.2 and 1.18 eV, FWHM = 43 and 44 meV, respectively.